\documentclass[prd,aps,12pt,showpacs,showkeys,nofootinbib]{revtex4-1}
\usepackage{graphicx}
\usepackage{setspace} 
\DeclareTextCommandDefault{\textcopyright}{\textcircled{c}}

\newcommand{\e}{\mbox{e}}

\begin{document}

\title{Acceleration \hspace*{-2.7mm} of  \hspace*{-2.7mm} cosmic  \hspace*{-2.7mm} expansion
  \hspace*{-2.7mm} through \hspace*{-2.7mm} huge \hspace*{-2.7mm} cosmological \hspace*{-2.7mm} constant
  \hspace*{-2.7mm} progressively \hspace*{-2.7mm} reduced \hspace*{-2.7mm} by \hspace*{-2.7mm} submicroscopic
  \hspace*{-2.7mm} information \hspace*{-2.7mm} transfer%
  \footnote{\mbox{Preprint of an article published in
      Int.~J.~Mod.~Phys.~A, Vol. 33, No. 23 (2018) 1850137 (21 pages),\\}
    \textcircled{c}~copyright~World~Scientific~Publishing~Company,\\
    DOI: 10.1142/S0217751X18501373}}

\author{E.~Rebhan\footnote{Eckhard.Rebhan@hhu.de}}
\address{Institut f{\"u}r Theoretische Physik,\\
   Heinrich--Heine--Universit{\"a}t,\\
   D--40225 D{\"u}sseldorf, Germany}

 \begin{abstract}
   \begin{singlespace}
   {\small
  In a previous paper (Ref.~\cite{Rebhan_1}) the presence of dark
  energy in our universe was explained as the fingerprint of a
  comprehensive, much older and expanding multiverse with positive
  spatial curvature, whose space-time is spanned by this energy, and
  which was created out of nothing. This concept is expanded by the
  addition of a model for explaining the decay of the mass density
  $\varrho$ of dark energy from its origin until now by a factor of
  approximately
  $10^{-120}$.\\
  Elementary particles contain information about which laws of nature
  they obey, but not what exactly these are. Most likely, the laws are
  not followed by obedience to a categorical imperative. Rather, it is
  assumed, that from the very beginning the information about them is
  coded in submicroscopic patches of the space-time. The initial
  density $\varrho_i$ is supposed to belong to the unimpaired
  cosmological constant obtained from elementary particle theory.  Due
  to its huge value it causes an extremely fast spatial expansion by
  which continuously new space-time elements are created. To them, the
  information about the physical laws must be transmitted from the
  already present space-time. This process needs time which with
  ever-increasing expansion velocity is getting scarcer and scarcer.
  It is concluded that this impedes the expansion through a
  friction-like process which can be described by a term proportional
  to the expansion-velocity. This term is subtracted from the
  expansion-acceleration. It is shown that the solutions thus obtained
  are also solutions of the cosmological standard equations employing
  a scalar field $\Phi$. In consequence, the present model can be
  considered as a re-interpretation of results which can be obtained
  with acknowledged methods.}
\end{singlespace}
 
\end{abstract}


\maketitle

\section{Introduction}
In this paper, the question is addressed how our universe manages,
that in each point of its space-time the physical laws are
obeyed. Elementary particles, the building bricks of matter, or the
fields representing them are endowed with specific properties which
enter the physical laws regulating their behavior. However, they do
not determine that fully but only in an indicative way by declaring
which laws apply to them. How they follow the laws will hardly
be virtual obedience to a categorical imperative. Examples from other
fields may serve as an illustration. 1.~In human society the laws to
be obeyed by the individuals are rooted in the brains of the
latter. Without the verifiable presence of this information they
cannot be followed. 2.~In biology, molecules that work together to
form a plant or an animal obtain the information about what to do by
the genetic code written down in the DNA. In both cases, the
information needed is physically present in all places where laws
control processes.

It is an empirical finding that once nature has developed a successful
recipe, it will employ it not only once but repeatedly. Therefore, it
appears natural to assume, that the information about the physical
laws to be followed by all material elements must somehow be contained
in the neighborhood of each space-time point. For the way how this is
accomplished, mainly two methods offer themselves: 1.~The information
is encoded in higher dimensions which are inaccessible to us but
connected with each spot of our space-time.  2. The information
entered the universe in the process of a creation out of nothing and
is encoded in the primordial space-time emerging from this. Both
methods are pretty well hidden, what they should actually be, because
so far they have not been discovered. In our view, both methods entail
the same consequences, and both of them can prove right or wrong. (Due
to its particular straightforwardness at some instances the second
possibility will be preferred.) However, the results derived from them
have attractive properties which are classified and rated in detail
later on.

As can be read from the title, in this paper, the process of
information transfer gets closely linked to the cosmological constant
problem \cite{Weinberg_1}. The latter was even the starting point for
this investigation. Indirectly this has turned up already in
Ref.~\cite{Rebhan_1} whose ideas are further developed here. Before
that, however, the most important results from there are briefly
recapitulated.

The question addressed in the previous paper is: why does our universe
contain dark energy, in view of the fact, that universes similar to
ours could well do without it. The answer given there is: our universe
is a subuniverse in an all-embracing and continuously expanding
multiverse, whose space-time is established by dark energy that also
fills our universe. The dark energy currently observed in the latter
thus represents a fingerprint of the multiverse containing it.

Measurements, from which the spatial curvature of our universe can be
deduced, render a value so small, that the space is either completely
uncurved and infinitely extended -- we refrain from topologically more
complex constellations like an uncurved space of finite extent --, or,
in the case of non-vanishing positive curvature, that it extends far
beyond the boundary of the observable universe \cite{Hinshaw, Planck}.
(The latter is in line with Ref.~\cite{Siegel} and with NASA's
conclusion from WMAP measurements \cite{NASA}: ``We now know (as of
2013) that the universe is flat with only a 0.4 \% margin of
error. This suggests that the Universe is infinite in extent; however,
.... All we can truly conclude is that the Universe is much larger
than the volume we can directly observe.'') In both cases there is a
large region outside the latter, most of which is not causally
connected with it and therefore structurally quite different from
it. This shows, that assuming the existence of a multiverse containing
our universe is justified, although there is no direct proof for
it. From the two above mentioned possibilities a positively curved
(and thus finitely extended) multiverse was chosen for several
reasons: 1. Only a positively curved multiverse can be created out of
nothing, which is of crucial importance for the purposes of this
paper. 2. Based on results from Ref.~\cite{Rebhan_2} it was shown
in Ref.~\cite{Rebhan_1} that the expansion of a positively curved
space has generic properties which do not pertain to an infinitely
extended uncurved space.

In a creation out of nothing, the multiverse emerges continuously from
a quantum mechanical tunneling process. For this, its expansion must
start at zero velocity immediately after the latter, and the initial
mass density~$\varrho_{\Lambda}$ of dark energy is fixed to about
$10^{120}$ times the value~$\varrho_0$ measured presently in our
universe \cite{Rebhan_1}. In Ref.~\cite{Rebhan_1} solutions of the
cosmological equations for a multiverse with dark energy as the
primary ingredient were derived. They have the required property that
the mass density $\varrho$ of the dark energy decays from a huge
initial value to an about $10^{-120}$ times smaller present value. Our
universe is one among other subuniverses, which are supposed to emerge
in the multiverse due to fluctuations of separate inflation fields of
their own, all of them permeated by the dark energy field of the
multiverse.  In properly chosen coordinates -- in particular with
identical proper time intervals $dt$ -- the mass density of the
latter, measured in our universe, is the same as measured in the
multiverse. The above factor $10^{-120}$ agrees more or less with the
factor by which a cosmological constant $\Lambda_0$ with the presently
observed mass density~$\varrho_0$ is smaller than predicted by
elementary particle theory \cite{Wiki}. This suggests that there is a
physical process which with increasing expansion of the multiverse
increasingly reduces the effect of $\varrho_\Lambda$. Complementary to
the findings of Ref.~\cite{Rebhan_1} in this paper a possible
mechanism for this is proposed and elaborated with respect to its
consequences.

Technically, in place of a huge cosmological constant $\Lambda$ we
employ the corresponding mass density $\varrho_{\Lambda}$ and consider
it as a contribution to the energy momentum tensor on the right hand
side of the Einstein equations, since conceptually it arises from
fluctuations of matter fields. We assume that it keeps its large
initial value during the whole evolution of the multiverse.  Without
an additional reduction process it would cause an extremely
accelerated expansion of the multiverse.  The reduction process
employed in this paper is the following: Firstly, we assume that the
information about all physical laws and the physical agents to whom
they relate, is transmitted through the creational tunneling process
to the subsequent initial state of the multiverse. Within the latter,
the information about the material agents is contained in these
themselves. Concerning the physical laws and their practical
implementation, we assume, borrowing from the biological DNA coding,
that the information about them is somehow coded in all places of the
primordial space-time within patches of submicroscopic but finite
extent.%
\footnote{For the storage of information in physical storage media,
  given their mass or energy and the amount of information, there is a
  size limit called Bekenstein bound~\cite{Bekenstein_1}, which can
  not be fallen below. In our case, while the space-time patches used
  for storage contain dark energy, it is not clear whether or not the
  latter is involved in the storage of information (later both options
  are considered). Furthermore, it is unclear whether the physics used
  to derive said boundary is applicable here at all.  Finally, for its
  application one would also have to know the amount of information to
  be stored. Here, this not only concerns the information contained in
  all physical laws, but also the definition of the quantities
  contained therein, the mathematics used in them, the logic
  underlying the latter, and in addition also the methods for
  practical implementation. Furthermore, it should be ensured that the
  physical laws known today are complete. An order of magnitude
  estimate of this amount of information seems feasible but would go
  far beyond the scope of this investigation. As it turns out, it is
  not necessary for the essential results of this work.}
Details about their structure (discrete or quantized \cite{Snyder, Guo,
  Ma} respectively, a quantum foam \cite{Chandra, NASA_2, Lobo},
perhaps even a structure below the quantum level, or
otherwise~\cite{Oriti}) and about the kind of encoding (DNA-like,
geometrically or otherwise) are not needed for the following and are
therefore not discussed. The additional space created by the
$\varrho_\Lambda$-induced spatial expansion must be equipped with the
information about the physical laws and physical properties of its own
(e.g. vacuum fluctuations) via transfer from already exis-
\newpage

\noindent
ting space. It is plausible that this process takes some time which is not
available, if the spatial expansion proceeds too fast.%
\footnote{This assumption is supported by the existence of an upper
  speed limit of information processing found by
  Bremermann~\cite{Bremermann,Bekenstein_2}, which is
  $1.36\cdot 10^{50}$ bits/(kg~s).}%
~In this regard, we now assume, that for gaining time the transfer of
information to the newly created patches of space-time impedes the
spatial expansion, and that this can be cumulatively represented by
introducing an appropriate friction term into the cosmological
equations. Without it, the relevant cosmological equation would be%
\vspace*{-0.5\baselineskip}
\begin{equation}
  \label{eq:1}
  \dot{a}^2(t) = \frac{8\pi G}{3}\,\varrho_\Lambda a^2 -c^2
\end{equation}
with the immediate consequence%
\vspace*{-0.4\baselineskip}
\begin{equation}
  \label{eq:2}
  \ddot{a}(t) = \frac{8\pi G}{3}\,\varrho_\Lambda a\,.
\end{equation}
The simplest way to include a friction term is a modification of this
equation in the form%
\vspace*{-0.5\baselineskip}
\begin{equation}
  \label{eq:3}
  \ddot{a}(t) = -f\dot{a}(t) +\frac{8\pi G}{3}\,\varrho_\Lambda a
\end{equation}
\vskip-0.5\baselineskip
\noindent
where $f{>\,}0$ is a constant. Multiplying this equation with
$\dot{a}(t)$ and integrating it with respect to $t$ yields
\begin{equation}
  \label{eq:4}
  \dot{a}^2(t) = \frac{8\pi G}{3}\varrho(t)\, a^2 -c^2
  \qquad\mbox{with}\qquad
  \varrho = \varrho_\Lambda + \varrho_f
\end{equation}
where
\vspace*{-\baselineskip}
\begin{equation}
  \label{eq:5}
 \varrho_f = - \frac{3\,f}{4\pi G\,a^2}\,\int_0^t \dot{a}^2(t')\,dt'
\end{equation}
and where an integration constant was chosen such that at the time
$t{=}0$ immediately after the tunneling process, the multiverse starts
with the unaltered mass density $ \varrho{=} \varrho_\Lambda$ and
assumes the form of the usual Friedman-Lemaitre equation (for short
\mbox{FL-equation}).  The result $\varrho_f{<\,}0$ means, that the
force field represented by $\varrho_f$ or $f\dot{a}(t)$ respectively
and introduced to reduce the effect of $\varrho_\Lambda$ via friction,
has negative mass or energy like the gravitational field. It therefore
makes sense to interpret it as a component of the gravitational field
described by the left side of the cosmological field equations. In
paragraph~(\ref{su:5}) of Section~\ref{sec:suppl} it is shown that the
density decomposition of Eq.~(\ref{eq:4}) with (\ref{eq:5}), caused by
the introduction of the friction term ${-}f\dot{a}(t)$ in
Eq.~(\ref{eq:3}), can be covariantly integrated into the Einstein
field equations.

Introducing an additional term in the cosmological equations is a
delicate matter that could easily meet with rejection.%
\footnote{Modifications of the Einstein field equations have indeed a
  long-standing tradition, beginning with Einstein's addition of the
  cosmological constant for obtaining a static universe (see e.g. page
  613 of Ref.~\cite{Weinberg_2}) to the introduction of additional
  mass generation terms for a steady state universe by H.~Bondi and
  T.~Gold (page 459 of Ref.~\cite{Weinberg_2}) and independently by
  F.~Hoyle (page 616 of Ref.~\cite{Weinberg_2}) up to very recent
  alterations as in Refs.~\cite{Novikov}-\cite{Abdel}.}
However, choosing the solutions of the modified equation~(\ref{eq:3})
such, that at all times the usual FL-equation~(\ref{eq:4}) holds,
leads back to the cosmological standard theory in which it suffices to
solve the latter. Therefore, it can be said that the newly introduced
model leads to standard solutions which are only re-interpreted in an
unusual way. In Sec.~\ref{sec:Interpret} it is shown that even the
cosmological equations for a scalar field $\Phi$, driven by a
potential $V(\Phi)$, are satisfied by our solutions.

Because of the information about particle-properties contained
therein, actually at least a cumulative mass density $\varrho_m$ for
matter should be included. However, for the sake of clarity, and since
it would not disclose anything new compared to the results of
Ref.~\cite{Rebhan_1}, we refrain from this and come back to it just
briefly in Sec.~\ref{sec:suppl}.

\section{Basic Equations and boundary conditions}
\label{sec:baseq}

All calculations are carried out in Friedman Robertson Walker
coordinates. In order to make numerical evaluations more transparent,
we use MSI units. The basic equations to be satisfied in a closed
multiverse with positive spatial curvature are
Eqs.~(\ref{eq:3})-(\ref{eq:4}), the dependent variables to be
calculated being $a(t)$ and $\varrho(t)$. Since Eq.~(\ref{eq:3})
contains only $a(t)$, this variable is already completely determined
by it, if appropriate boundary conditions are imposed. As soon as
$a(t)$ is known, Eq.~(\ref{eq:4}) can be used for calculating
$\varrho(t)$.  For determining $\varrho_f(t)$, Eq.~(\ref{eq:4}) is not
even needed since it can be calculated from
$\varrho_f{=}\varrho(t){-}\varrho_\Lambda$. For later purposes we
still need the equations%
\vspace*{-0.5\baselineskip}
\begin{eqnarray}
      &&  \qquad \varrho =  \frac{\hbar^2 \dot{\Phi}^2(t)}{2\mu c^4}+ \frac{V(\Phi)}{c^2}\,,
                  \quad\quad
      p = \frac{\hbar^2 \dot{\Phi}^2(t)}{2\mu c^2} - V(\Phi)\,,
  \label{eq:1.1}\\   
          &&\qquad  \ddot{\Phi}(t) + 3(\dot{a}(t)/a)\,\dot{\Phi}(t)
             + \frac{\mu c^2}{\hbar^2}\,V'(\Phi) = 0
  \label{eq:1.2}              
\end{eqnarray}
\vskip-0.5\baselineskip
\noindent
which hold when the expansion of the multiverse is driven by a scalar
field $\Phi$ ($\mu$ is the mass parameter of the field~$\Phi$).

As already mentioned in the Introduction, for $t{=}0$ the
initial conditions
\begin{equation}
  \label{eq:1.7}
  \varrho = \varrho_\Lambda\,,\qquad a = a_i = l_P\,,\qquad \dot{a}(t) = 0
\end{equation}
must be satisfied. With this and $l_P/t_P{=}c$, from Eq.~(\ref{eq:4})
we obtain
\begin{equation}
  \label{eq:1.8}
  \varrho_\Lambda = \frac{3c^2}{8\pi G\,a_i^2}
    \qquad\mbox{or}\qquad
    \frac{8\pi Gt_P^2}{3} = \frac{1}{\varrho_\Lambda}
\end{equation}
and
\begin{equation}
  \label{eq:1.9}
  \varrho_\Lambda =   \frac{3}{8\pi G}\,\varrho_P
  = 0.616\cdot 10^{96}\,\mbox{kg m}^{-3}\,,
\end{equation}
where $l_P$ and $t_P$ are given immediately below and
$\varrho_P{=}c^5/(\hbar G^2)$ is the Planck density. A further
boundary condition concerning the mass density $\varrho_0$ at the
present time $t_0$ is dealt with in Sec.~\ref{sec:varrho}.

For the evaluation of Eqs.~(\ref{eq:3})-(\ref{eq:1.2}) it is useful to
introduce dimensionless quantities
\begin{eqnarray}  
  \tau &=& \frac{t}{t_P}
  \qquad\mbox{with}\qquad
  t_P  = \sqrt{\frac{\hbar G}{c^5}}=5.39\cdot 10^{-44}\,\mbox{s}
  \label{eq:1.3}\\
  x &=& \frac{a}{l_P}
  \qquad\mbox{with}\qquad\;
  l_{P} = \sqrt{\frac{\hbar G}{c^3}} = 1.616\cdot 10^{-35}\,\mbox{m}\,,
  \label{eq:1.4}\\
  \rho &=& \frac{\varrho}{\varrho_{\Lambda}}\,,\hspace*{19.5mm}
           \hat{\pi} = \frac{p}{\varrho_{\Lambda} c^2}\,,
  \label{eq:1.5}\\
  \varphi  &=& \frac{\hbar}{c^2}\sqrt{\frac{8\pi G}{3\mu}}\,\Phi\,, \!\quad\quad
               v(\varphi) = \frac{V(\Phi)}{\varrho_{\Lambda} c^2}\,,
  \label{eq:1.6}
\end{eqnarray}
where $t_P$ and $l_P$ are the Planck time and the Planck length
respectively.  Using Eqs.~(\ref{eq:1.8}) and
(\ref{eq:1.3})-(\ref{eq:1.5}), Eq.~(\ref{eq:4}) becomes
\vspace*{-0.5\baselineskip}
\begin{equation}
  \label{eq:1.10}
  \dot{x}^2(\tau) = \rho\,x^2 - 1 \,, 
\end{equation}
and using Eqs.~(\ref{eq:1.6}) in addition,
Eqs.~(\ref{eq:1.1})-(\ref{eq:1.2}) become
\begin{equation}
  \label{eq:1.11}
  \rho = \frac{\dot{\varphi}^2(\tau)}{2} + v(\varphi) \,, \qquad
  \hat{\pi} = \frac{\dot{\varphi}^2(\tau)}{2} - v(\varphi) 
\end{equation}
and
\vspace*{-0.8\baselineskip}
\begin{equation}
  \label{eq:1.12}
  \ddot{\varphi}(\tau)+3\,h\,\dot{\varphi}(\tau)+v'(\varphi) = 0
  \quad\mbox{with}\quad
  h = \frac{\dot{x}(\tau)}{x}\,.
\end{equation}
Finally, Eq.~(\ref{eq:3}) becomes
\vspace*{-0.5\baselineskip}
\begin{equation}
  \label{eq:1.13}
  \ddot{x}(\tau) + 2\sigma\dot{x}(\tau) -x = 0
  \quad\mbox{with}\quad
  \sigma = \frac{f\,t_P}{2}\,.
\end{equation}
\vskip-0.5\baselineskip
\noindent
This equation must be solved with appropriate boundary conditions. In
terms of the dimensionless variables~(\ref{eq:1.3})-(\ref{eq:1.4}),
from the initial conditions~(\ref{eq:1.7}) the relevant ones become%
\vspace*{-0.5\baselineskip}
\begin{equation}
  \label{eq:1.14}
  x = 1\,,\qquad \dot{x}(\tau) = 0
  \qquad\mbox{for}\quad \tau =0\,.
\end{equation}
\vskip-0.8\baselineskip
\noindent
The mass density corresponding to the solution $x(\tau)$ follows from
Eq.~(\ref{eq:1.10}) and is
\begin{equation}
  \label{eq:1.15}
  \rho(\tau) = \frac{1}{x^2(\tau)} + \frac{\dot{x}^2(\tau)}{x^2(\tau)}\,.
\end{equation}
To find out how the solutions of Eq.~(\ref{eq:1.13}) compare to
solutions of the usual scalar field theory, we express the equation%
\vspace*{-0.5\baselineskip}
\begin{displaymath}
  \ddot{a}(t) = -\frac{4\pi G\,\varrho}{3}\,(1+3w)\,a
  \qquad\mbox{with}\qquad
  w = \frac{p}{\varrho\,c^2} \,,
\end{displaymath}
obtained with use of Eqs.~(\ref{eq:1.1})-(\ref{eq:1.2}) from the time
derivative of Eq.~(\ref{eq:4}), in terms of the dimensionless
variables~(\ref{eq:1.4})-(\ref{eq:1.5}) and obtain
\vspace*{-0.5\baselineskip}
\begin{equation}
  \label{eq:1.16}
   \ddot{x}(\tau) = -\frac{1+3w}{2}\,\rho\,x\,.
 \end{equation}
 
\section{Solutions}
\label{sec:Solutions}

\subsection{Calculation of the expansion $x(\tau)$}

Eq.~(\ref{eq:1.13}) can be solved with the ansatz
$x(\tau){=}\e^{\gamma \tau}$. The general solution thus obtained is
\vspace*{-0.5\baselineskip}
\begin{equation}
  \label{eq:1.17}
  x(\tau) = \alpha\,\e^{\gamma_1 \tau} + \beta \,\e^{\gamma_2 \tau}
\end{equation}
\vskip-0.8\baselineskip
\noindent
with
\vspace*{-0.2\baselineskip}
\begin{equation}
  \label{eq:1.18}
  \gamma_1 = \sqrt{1{+}\sigma^2}-\sigma\,, \qquad
  \gamma_2 = \sqrt{1{+}\sigma^2}+\sigma\,.
\end{equation}
Imposing on it the initial conditions~(\ref{eq:1.14}) yields
\begin{displaymath}
  \alpha+\beta = 1\,, \qquad  \alpha\gamma_1  + \beta\gamma_2 = 0\,.
\end{displaymath}
From this, with use of Eqs.~(\ref{eq:1.18}) we get
\begin{equation}
  \label{eq:1.19}
  \alpha = \frac{\gamma_2}{2\,\sqrt{1{+}\sigma^2}} = \frac{1}{1{+}\gamma^2}\,,\qquad
  \beta  = \frac{\gamma_1}{2\,\sqrt{1{+}\sigma^2}} = \frac{\gamma^2}{1{+}\gamma^2}\,,
\end{equation}
where the relations $\gamma_1\gamma_2{=}1$ and
$\gamma_1{+}\gamma_2{=}2\,\sqrt{1{+}\sigma^2}$, following from
Eqs.~(\ref{eq:1.18}), were used and the redefinition%
\vspace*{-0.5\baselineskip}
\begin{equation}
  \label{eq:1.21}
  \gamma =  \gamma_1 = \sqrt{1{+}\sigma^2}-\sigma
\end{equation}
was carried out. Inserting the results~(\ref{eq:1.19}) in
Eq.~(\ref{eq:1.17}) leads to
\begin{equation}
  \label{eq:1.22}
  x(\tau) = \frac{\e^{\gamma\tau}{+}\gamma^2\e^{-\tau/\gamma}}{1+\gamma^2}
\end{equation}
as the final form of our solution. Resolving Eq.~(\ref{eq:1.21}) with
respect to $\sigma$ yields
\begin{equation}
  \label{eq:1.23}
  \sigma = \frac{1{-}\gamma^2}{2\,\gamma}\,.
\end{equation}
The result~(\ref{eq:1.22}) can be rewritten in the form
\begin{displaymath}
  x(\tau) = x_1(\tau) + x_2(\tau)
  \qquad\mbox{with}\qquad
  x_1(\tau) =  \e^{\gamma\tau}\,,\qquad
  x_2(\tau) = -\frac{\gamma^2}{1{+}\gamma^2}(\e^{\gamma\tau}{-}\e^{-\tau/\gamma})\,.
\end{displaymath}
For the ratio $q{=}|x_2(\tau)/x_1(\tau)|$ we get
\begin{displaymath}
  q = \frac{\gamma^2}{1{+}\gamma^2}(1{-} \e^{-(\gamma+1/\gamma)\tau})
  \leq \frac{\gamma^2}{1{+}\gamma^2} < \gamma^2 = 1.02\cdot 10^{-122}\,,
\end{displaymath}
where in the last step the later result~(\ref{eq:1.32}) was
introduced. It follows that with just an extremely small
error we can put
\vspace*{-0.5\baselineskip}
\begin{equation}
  \label{eq:1.24}
  x(\tau) = \e^{\gamma\tau}
  \qquad\mbox{for~all}\quad
  \tau \geq 0\,.
\end{equation}
Note that by this approximation the first of the
conditions~(\ref{eq:1.14}) is satisfied exactly and the second
approximately, but due to $\gamma{\ll}1$ with an extremely small error
only.

\subsection{Calculation of the mass density  $\varrho$
  and determination of $\gamma$}
\label{sec:varrho}

As in Ref.~\cite{Rebhan_1} we assume that the agent driving the
expansion of the multiverse is also responsible for the acceleration
of the expansion presently observed in our universe. There, it was
shown that the mass density of the corresponding dark energy in our
universe can be identified directly with the mass density $\varrho$ of
the expansion field in the multiverse. Accordingly, besides the
initial conditions~(\ref{eq:1.7}) or (\ref{eq:1.14}) respectively, at
a much later point of time $t_0$, representing the present age of our
universe in terms of the time $t$ measured in the multiverse, the
further boundary condition
\begin{equation}
  \label{eq:1.25}
  \varrho_0 = \varrho(t_0) = 0,683\,\varrho_{c0}
  \qquad\mbox{with}\quad
  \varrho_{c0} = \frac{3H_0^2}{8\pi G} 
  = 9.20\cdot 10^{-27}\,\mbox{kg\,m}^{-3}
\end{equation}
must be met. Thereby $\varrho_{c0}$ is the critical density and $H_0$
the present Hubble parameter of our universe. Furthermore, for the
important ratio $\varrho_0/\varrho_\Lambda$ we obtain with
Eq.~(\ref{eq:1.9}) the more accurate value%
\vspace*{-0.2\baselineskip}
\begin{equation}
  \label{eq:1.25+}
  \frac{\varrho_0}{\varrho_\Lambda} = 1.02\cdot 10^{-122}\,.
\end{equation}
For the evaluation of the boundary condition it is recommended to
employ a different subset of dimensionless variables which is better
adjusted to the macroscopic scales involved, namely%
\vspace*{-0.2\baselineskip}
\begin{equation}
  \label{eq:1.26}
  X = \frac{a}{a_0} = \frac{a}{R\zeta}
  \qquad\mbox{with}\quad
  \zeta =  \frac{a_0}{R}\,,
\end{equation}
\vskip-1.45\baselineskip
\noindent
where
\vspace*{-0.3\baselineskip}
\begin{equation}
  \label{eq:1.27}
  R = a_0\,r_{bo} = 23.5\cdot 10^{9}\,\mbox{ly} = 2.22\cdot 10^{26}\,\mbox{m}
\end{equation}
is the present metric radius of the boundary of our observable
universe (at radial coordinate $r_{bo}$ and for scale factor
$a_0{=}a(t_0)$), and
\begin{equation}
  \label{eq:1.28}
  \mathcal{T} = \frac{t}{t_{H_0}}
    \qquad\mbox{with}\quad
    t_{H0} = \sqrt{\frac{3}{8\pi G \varrho_{c0}}} = \frac{1}{H_0} 
  = 14.0\cdot 10^{9}\,\mbox{a}  = 4.41\cdot 10^{17}\,\mbox{s} \,,
\end{equation}
where $ t_{H0}$ is the present Hubble time.

Now we calculate the density $\varrho_0{=}\varrho(t_0)$, employing the
approximation~(\ref{eq:1.24}) for $x(\tau)$. With this and its
consequence $\dot{x}(\tau){=}\gamma\,x$, from Eq.~(\ref{eq:1.15}) we
get
\vspace*{-0.5\baselineskip}
\begin{equation}
  \label{eq:1.29}
  \varrho = \varrho_\Lambda\gamma^2 + \frac{\varrho_\Lambda}{x^2}\,.
\end{equation}
\vskip-1.5\baselineskip
\noindent
With the relations
\begin{equation}
  \label{eq:1.30}
  \tau = \frac{ t_{H0}}{t_P}\, \mathcal{T}\,, \qquad
  x = \frac{R\zeta X}{l_P}
\end{equation}
following from Eqs.~(\ref{eq:1.3})-(\ref{eq:1.4}), (\ref{eq:1.26}) and
(\ref{eq:1.28}), and inserting numbers by using Eqs.~(\ref{eq:1.9}),
(\ref{eq:1.4}), (\ref{eq:1.25}) as well as Eq.~(\ref{eq:1.27}),
Eq.~(\ref{eq:1.29}) becomes
\begin{equation}
  \label{eq:1.31}
  \frac{\varrho}{\varrho_0} =  \frac{\varrho_\Lambda}{\varrho_0}\,\gamma^2
  + \frac{0.522}{\zeta^2X^2}\,.
\end{equation}
(Note that in the calculation of the second term, the exponents of the
extremely large value
$\varrho_\Lambda/\varrho_0{=}0.98{\cdot}10^{122}$ and the extremely
small value $(l_P/R)^2{=}0.53{\cdot}10^{-122}$ cancel each other out
quite naturally.) From this, for $\mathcal{T}{=}\mathcal{T}_0$ with
$X(\mathcal{T}_0){=}1$ according to Eq.~(\ref{eq:1.26}), and
restricting ourselves to sufficiently large values
$\zeta{\gtrsim}10^2$ (such that $1/\zeta^2{\lesssim}10^{-4}$), we
obtain
\begin{equation}
  \label{eq:1.32}
  \gamma = \sqrt{\frac{\varrho_0}{\varrho_\Lambda}} = 1.01\cdot 10^{-61}\,.
\end{equation}
Inserting this result in Eqs.~(\ref{eq:1.29}) and (\ref{eq:1.31})
yields
\begin{equation}
  \label{eq:1.33}
  \varrho = \varrho_0 + \frac{\varrho_\Lambda}{x^2}
  = \left(1+\frac{0.522}{\zeta^2X^2}\right)\varrho_0\,.
\end{equation}
It follows immediately that $\varrho'(x){<\,}0$, a condition needed in
Sec.~\ref{sec:Interpret}. However, we have to investigate still more
closely the immediate neighborhood of $x{=}1$ or $\tau{=}0$
respectively, since the approximation~(\ref{eq:1.24}), used for
deriving Eq.~(\ref{eq:1.29}), does not satisfy
$\dot{x}(\tau)|_{\tau=0}{=}0$ as required. For this purpose, we must
go back to the exact equation~(\ref{eq:1.15}), and insert in it
$\dot{x}(\tau)$ as obtained from Eq.~(\ref{eq:1.22}), getting
\begin{equation}
  \label{eq:1.34}
  \frac{\varrho}{\varrho_\Lambda} 
  =  \frac{1}{x^2} + \gamma^2 \left(\frac{1{-}\e^{-\tau(\gamma+1/\gamma)}}
    {1{+}\gamma^2\e^{-\tau(\gamma+1/\gamma)}}\right)^2
  =\frac{1}{x^2} + \mathcal{O}(10^{-122})\,.
\end{equation}
Accordingly, we have with high precision
$\varrho'(x){=}{-}2\varrho_\Lambda/x^3{<\,}0$ at and around $x{=}1$.

Using Eqs.~(\ref{eq:1.5}a),%
\footnote{ Eq.~(\ref{eq:1.5}a) denotes the first of the
  Eqs.~(\ref{eq:1.5}), Eq.~(\ref{eq:1.5}b) the second, etc.}%
~(\ref{eq:1.24}) and (\ref{eq:1.29}), the resolution of
Eq.~(\ref{eq:1.16}) with respect to $w$ yields%
\begin{equation}
  \label{eq:1.36}
  w = -1 + \frac{2}{3\,(1{+}\gamma^2x^2)} =
  -1 + \frac{2}{3\,(1{+}1.92\,\zeta^2X^2)}\,,
\end{equation}
where in the last step Eq.~(\ref{eq:1.30}b) and
$(\gamma R/l_P)^2{=}1.92$ according to Eqs.~(\ref{eq:1.4}),
(\ref{eq:1.27}) and (\ref{eq:1.32}) was used.

\subsection{Relation between $\mathcal{T}$ and $\zeta$}

Using Eqs.~(\ref{eq:1.3}), (\ref{eq:1.4}), (\ref{eq:1.27}),
(\ref{eq:1.28}) and (\ref{eq:1.32}), we obtain from
Eqs.~(\ref{eq:1.30}) for $\zeta{\gtrsim}10^2$
\begin{displaymath}
  \gamma\tau = 0.826\,\mathcal{T}
  \qquad\mbox{and}\qquad
  x = 1.37\cdot 10^{61}\,\zeta X\,.
\end{displaymath}
From this and Eq.~(\ref{eq:1.24}) we get
\begin{equation}
  \label{eq:1.37}
  X(\mathcal{T}) = \frac{\e^{0.826\,\mathcal{T}}}{1.37\,\zeta}\cdot 10^{-61}
  \qquad\mbox{or}\qquad
  \mathcal{T} = 170.35 + 1.21\,(\ln\zeta{+}\ln X)\,.
\end{equation}

\subsection{Tunneling solution for $t{<\,}0$}

We assumed that the mass density $\varrho_\Lambda$ emerges unreduced
from a creation out of nothing. The latter can be described
approximately by a tunneling solution, which means that for $t{\leq}0$
we can set $\varrho{\equiv}\varrho_\Lambda$ and must only solve
Eq.~(\ref{eq:1}) with the boundary condition $\dot{a}(t){=}0$ for
$t{=}0$. In the dimensionless variables~(\ref{eq:1.3})-(\ref{eq:1.4})
this equation becomes%
\begin{equation}
  \label{eq:1.38}
  \dot{x}^2(\tau) = x^2 - 1\,,
\end{equation}
and the boundary condition becomes $x{=}1$ for $\tau{=}0$. As in
Ref.~\cite{Rebhan_1}, for $\tau{\leq}0$ we set
\begin{equation}
  \label{eq:1.39}
  \tau = -\mbox{i}\,u
\end{equation}
with what  Eq.~(\ref{eq:1.38}) becomes
\vspace*{-0.1\baselineskip}
\begin{equation}
  \label{eq:1.40}
  \dot{x}^2(u) = 1-x^2\,. 
\end{equation}
The solution to the boundary condition $\dot{x}(u){=}0$ is
\begin{equation}
  \label{eq:1.41}
  x(u) = \cos u\,. 
\end{equation}

\subsection{Properties of the solutions}

\subsubsection{Solution $x(\tau)$}

\begin{figure}
  \includegraphics[trim=2cm 20cm 3cm 4cm, clip=true]{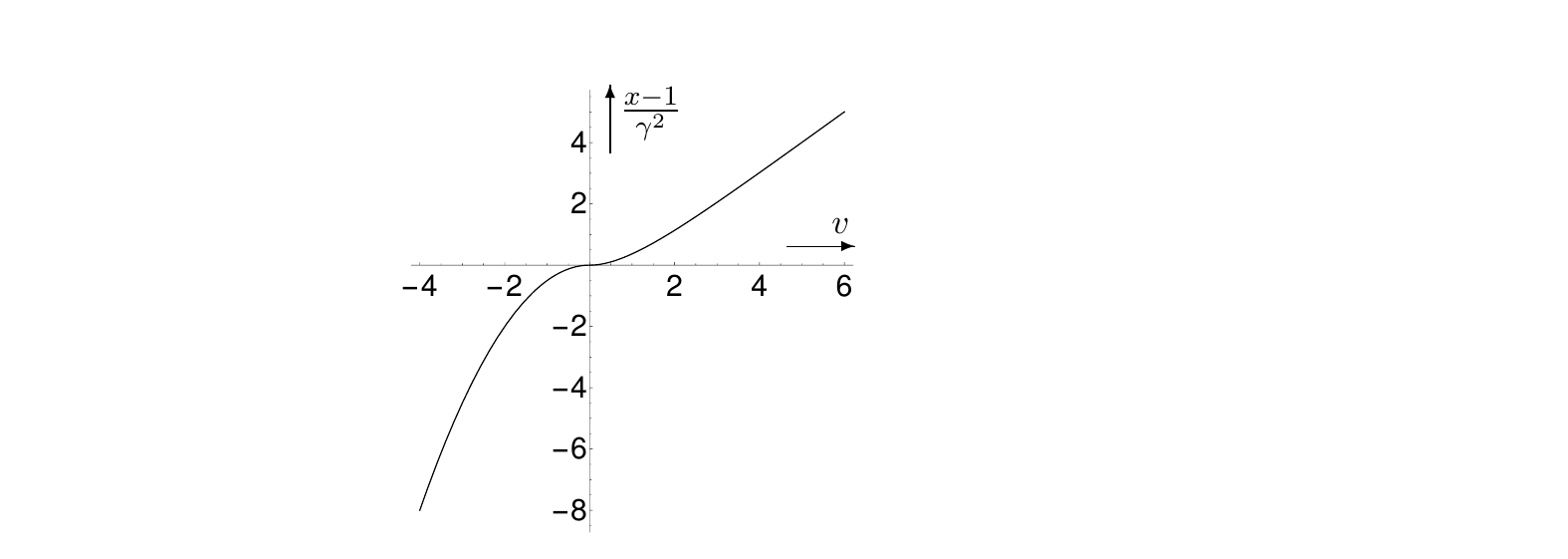} 
  \caption{\label{fig:1} $(x{-}1)/\gamma^2$ depending on
    $v{=}\tau/\gamma$ in the immediate neighborhood of $\tau{=}0$}
\end{figure}

First, we consider $x(\tau)$ in the immediate neighborhood of
$\tau{=}0$.  For $\tau{\geq}0$ we must use Eq.~(\ref{eq:1.22}) and for
$\tau{\leq}0$ Eq.~(\ref{eq:1.41}), in both cases $\gamma$ given by
Eq.~(\ref{eq:1.32}). Due to the extreme smallness of $\gamma$ it is
useful to introduce a new variable $v$ by
\begin{equation}
  \label{eq:1.42}
  \tau = \gamma v \quad\mbox{for}\quad v \geq 0\,,
  \qquad
  u = \gamma v  \quad\mbox{for}\quad v \leq 0\,.
\end{equation}
Expanding Eqs.~(\ref{eq:1.22}) and (\ref{eq:1.41}) with respect to
$\gamma^2$, with $\e^{-\tau/\gamma}{=\,}\e^{-v}$, 
\begin{displaymath}
  \e^{\gamma\tau} = 1 + \gamma^2v + \mathcal{O}(\gamma^4)
  \qquad\mbox{and}\qquad
  \cos(\gamma v) = 1 - \frac{\gamma^2v^2}{2} + \mathcal{O}(\gamma^4)
\end{displaymath}
we get
\vspace*{-0.5\baselineskip}
\begin{equation}
  \label{eq:1.43}
  \frac{x{-}1}{\gamma^2} = \left\{
    \begin{array}{cll}
      v + \e^{-v} -1 + \mathcal{O}(\gamma^4) &\quad\mbox{for}&\;v\geq 0\\
      -v^2/2 + \mathcal{O}(\gamma^4) &\quad\mbox{for}&\;v\leq 0\,.
    \end{array}
 \right.
\end{equation}
This result is represented graphically in FIG.~\ref{fig:1}.
\begin{figure}
  \includegraphics[trim=3cm 22cm 3cm 4cm, clip=true]{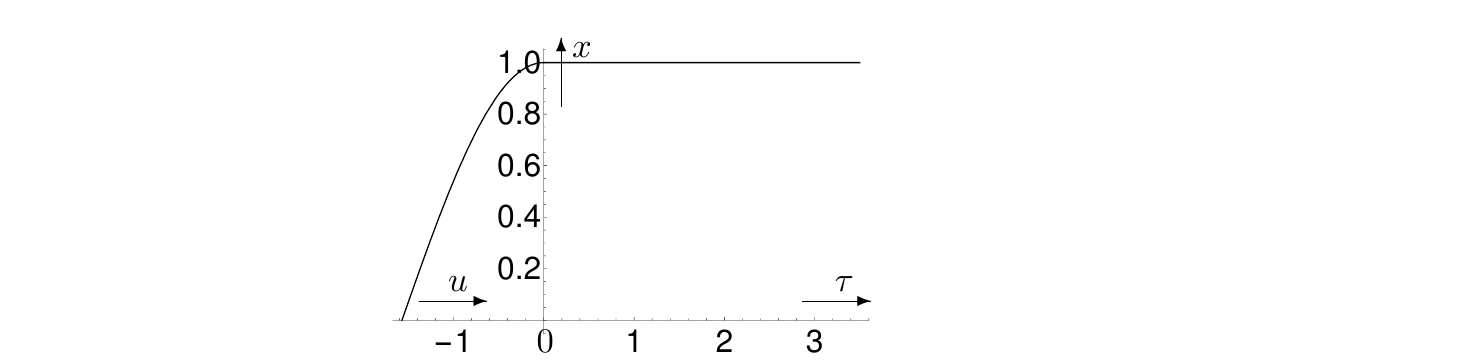} 
  \caption{\label{fig:2} $x(\tau)$ in the neighborhood of $\tau{=}0$}
\end{figure}

Plotting $x$ over $\tau$ or $u$ respectively gives a totally different
picture, which is shown in FIG.~\ref{fig:2}. According to
Eqs.~(\ref{eq:1.22}) and (\ref{eq:1.41}), we have
\begin{displaymath}
  x(u) = \left\{
    \begin{array}{llc}
      0 &\quad\mbox{for}&u=-\pi/2\\
      1 &\quad\mbox{for}& u=0
    \end{array}
  \right.\,,\qquad
   x(\tau) = \left\{
    \begin{array}{clc}
      1 &\quad\mbox{for}&\tau = 0\\
      1{+}\mathcal{O}(10^{-61})&\quad\mbox{for}&\tau = \pi/2\,,
    \end{array}
  \right.
\end{displaymath}
which means, that $x(\tau)$ remains extremely close to its initial
value $x{=}1$. In order to determine for how long this will be, we
calculate the time from when on $x$ deviates more than $1$ percent
from $x{=}1$. From Eqs.~(\ref{eq:1.3}), (\ref{eq:1.24}),
(\ref{eq:1.28}) and (\ref{eq:1.32}) we get
\begin{displaymath}
  \tau = \frac{\ln 1.01}{\gamma} = 0.985\cdot 10^{59}
  \qquad\mbox{or}\qquad
  t = 5.31\cdot 10^{15}\,\mbox{s} = 1,20\cdot 10^{-2}\,t_{H_0}
  = 1.69\cdot 10^8\mbox{y}\,.
\end{displaymath}
Only after this very long time $x$ deviates notably from $1$.

\begin{figure}
  \includegraphics[trim=3cm 21.9cm 3cm 4cm, clip=true]{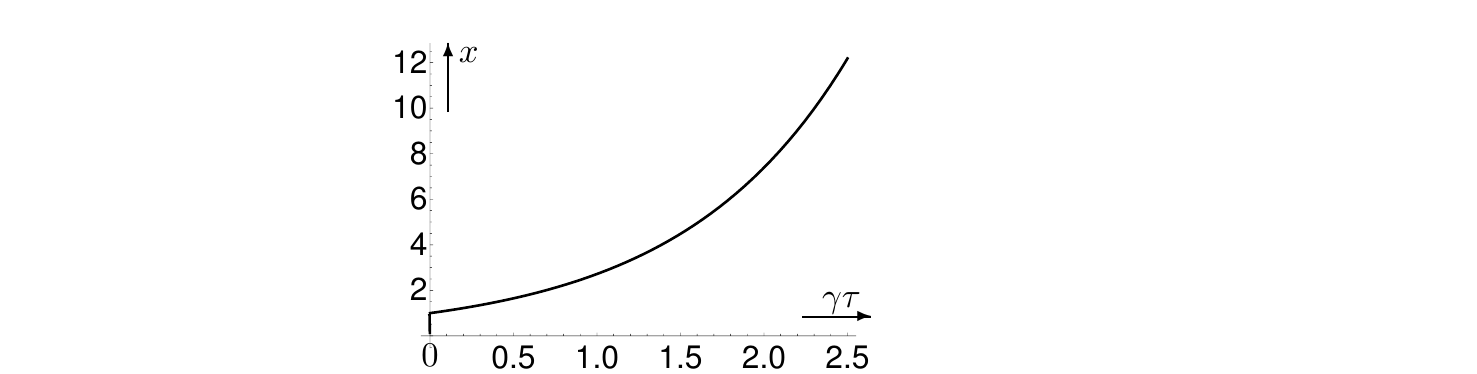} 
  \caption{\label{fig:3} $x(\tau)$ depending on
    $v{=}\gamma\tau{=}0.83\,\mathcal{T}$ in the wider neighborhood of
    $\tau{=}0$}
\end{figure}

By plotting $x$ over $v{=}\gamma\tau$ for $v{\geq}0$ and
$v{=}\gamma u$ for $v{\leq}0$, still another picture is obtained,
which is shown in FIG.~\ref{fig:3}. For $v{\leq}0$ we have
$ x(v){=}\cos(v/\gamma){\approx}\cos(10^{61}v)$ whence $x{=}0$ for
$v{\approx}10^{-61}\pi/2$ and $x{=}1$ for $v{=}0$, i.e. the interval
${-}\pi/2{\leq}u{\leq}0$ has shrunk to an invisible size. On the other
hand for $v{\geq}0$ we have $ x(v){=}\e^v$.

\subsubsection{Solutions $\mathcal{T}_0(\zeta)$, $\varrho(x)$ and $w(x)$}

Using $X{=}1$ for $\mathcal{T}{=}\mathcal{T}_0$, Eq.~(\ref{eq:1.37}) gives
\begin{equation}
  \label{eq:1.44}
  \mathcal{T}_0 = 170.35 +1.21\,\ln\zeta
  \qquad\mbox{and}\qquad
  \mathcal{T} = \mathcal{T}_0 + 1.21\,\ln X \,.
\end{equation}
For $\zeta{\geq}10^2$, required for reasons of simplicity in the
derivation of Eq.~(\ref{eq:1.32}), we obtain from this that
$\mathcal{T}_0{\geq}176$; furthermore, $\mathcal{T}_0{=}178.7$ for
$\zeta{=}10^3$ and $\mathcal{T}_0{=}181.5$ for $\zeta{=}10^4$. As in
the case of the solutions obtained in Ref.~\cite{Rebhan_1}, the
condition that our universe should fit into the multiverse both time-
and space-wise, can easily be satisfied.

\begin{figure}
  \includegraphics[trim=3cm 21.2cm 3cm 4cm, clip=true]{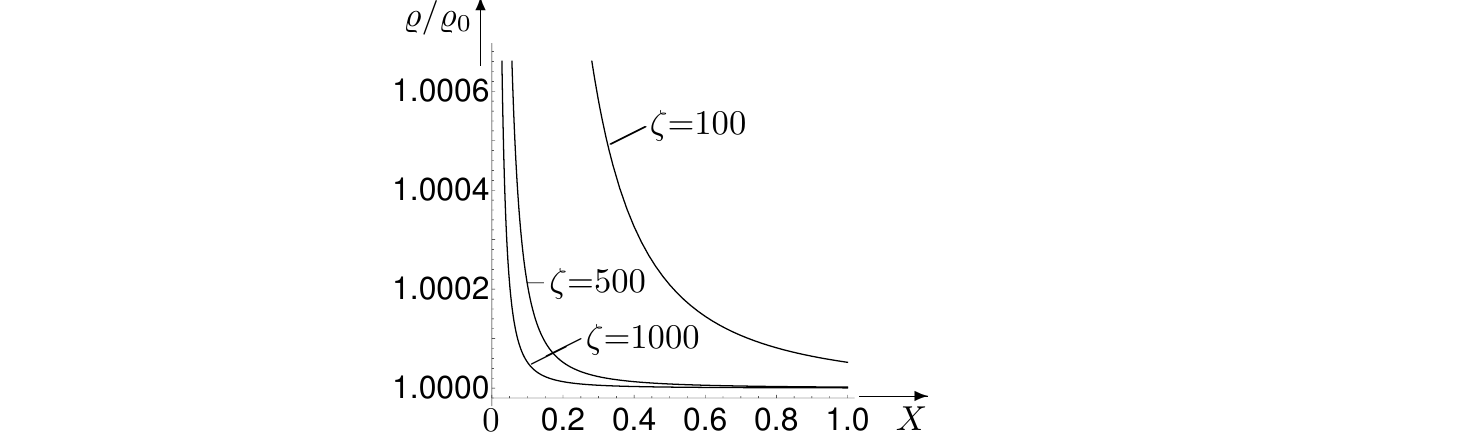} 
  \caption{\label{fig:4} Density ratio $\varrho/\varrho_0$ as a
    function of $X$ for several values of $\zeta$}
\end{figure}

For $x{\gtrsim}1$, the density $\varrho$ decays according to
Eq.~(\ref{eq:1.34}); for $x{\gg}1$, it decays according to
Eq.~(\ref{eq:1.33}) what is shown in FIG.~\ref{fig:4} for
$X{\lesssim}1$ and several values of~$\zeta$. It is seen that
$\varrho$ approaches its present value already for $X$-values
significantly below $1$. In particular we have
\begin{equation}
  \label{eq:1.45}
  \frac{\varrho}{\varrho_0}-1 \leq 10^{-2}
    \quad\mbox{for}\quad
    X \geq \left\{
      \begin{array}{lll}
        0.72\cdot 10^{-1} \;\mathrel{\widehat{=}}\;\mathcal{T}= \mathcal{T}_0{-}3.2
        \qquad&\mbox{when}&\quad \zeta=10^2\,,\\
        0.72\cdot 10^{-2}  \;\mathrel{\widehat{=}} \;\mathcal{T}=\mathcal{T}_0{-}6
        \qquad&\mbox{when}&\quad \zeta=10^3\,,
      \end{array}
\right.
\end{equation}
where at last Eqs.~(\ref{eq:1.44}) were used. During the lifetime of
our universe, i.e. for
$\mathcal{T}_0{-}0.98{\leq}\mathcal{T}{\leq}\mathcal{T}_0$, the
density $\varrho$ changes still much less, according to
Eqs.~(\ref{eq:1.33}) and (\ref{eq:1.37}) only by
$\Delta\varrho{\approx}10^{-4}\varrho_0$ for $\zeta{=}10^2$ and
$\Delta\varrho{\approx}10^{-6}\varrho_0$ for $\zeta{=}10^3$.
\begin{figure}
  \includegraphics[trim=3cm 22cm 3cm 3.9cm, clip=true]{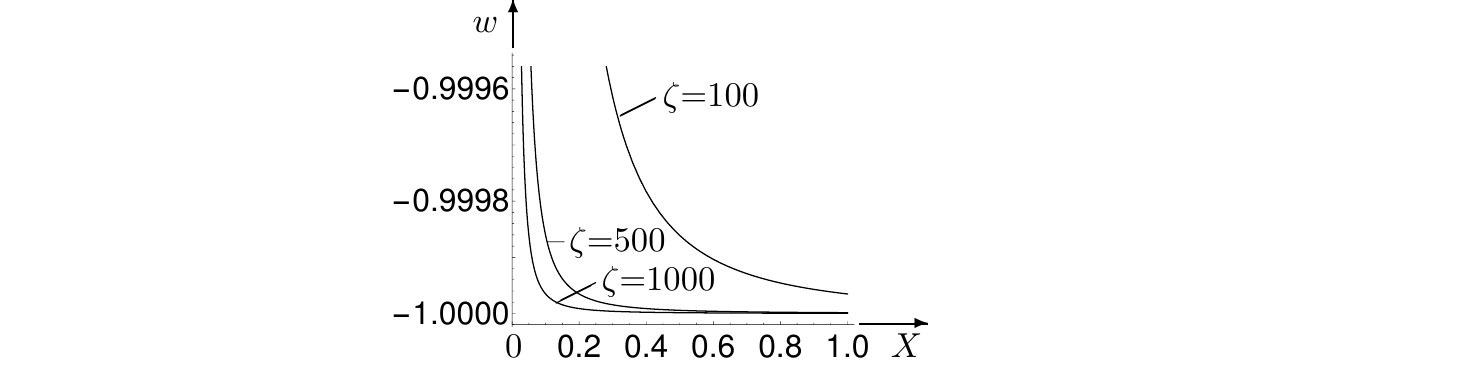} 
  \caption{\label{fig:5} Ratio $w{=}p/(\varrho c^2)$ as a
    function of $X$ for several values of $\zeta$}
\end{figure}

According to Eqs.~(\ref{eq:1.36}), $w$ starts essentially with
$w{=}{-}1/3$ at $x{=}1$ and converges to $w{=}{-}1$ for
$x{\to}\infty$. The present value, obtained for $X{=}1$, is given by
\begin{displaymath}
   w_0 =  -1 + \frac{2}{3\,(1{+}1.92\,\zeta^2)}\,.
\end{displaymath}
From this follows%
\vspace*{-0.5\baselineskip}
\begin{displaymath}
  |w_0+1|\leq  \left\{
     \begin{array}{lll}
      3.5\cdot 10^{-5} &\quad\mbox{for}&\zeta=10^2\\
      3.5\cdot 10^{-7} &\quad\mbox{for}&\zeta=10^3\,,
    \end{array}
\right.
\end{displaymath}
i.e. up to a very small error we have $w_0{=}{-}1$ for all
$\zeta$-values considered (FIG.~\ref{fig:5}). This is essentially the
same value as the one obtained from the cosmological constant model,
which according to Ref.~\cite{Wang} has "so far the best performance
in fitting the observational data" (for data see
e.g. Ref.~\cite{Planck} or Ref.~\cite{Betoule}). According to
Eqs.~(\ref{eq:1.36}) and (\ref{eq:1.37}), as time goes by, our model
is approaching closer and closer to the cosmological constant model so
that it is virtually indistinguishable from it during the lifetime of
our universe.

\subsubsection{Curvature parameter $\Omega_k$}

As already stated in the Introduction, the spatial curvature measured
in our universe is very small. The parameter commonly used for its
representation (see e.g. p. 41 of Ref.~\cite{Weinberg_3} or p. 491
of Ref.~\cite{Rebhan_3}) is%
\vspace*{-0.7\baselineskip}
\begin{equation}
  \label{eq:1.44+}
  \Omega_k = -\frac{k\,c^2}{a_0^2\,H_0^2}
  = -\frac{c^2\,t^2_{H_0}}{R^2\,\zeta^2} = -\frac{0.36}{\zeta^2}\,, 
\end{equation}
where in the last two steps $X_0{=}1$ and
Eqs.~(\ref{eq:1.26})-(\ref{eq:1.28}) were used. According to the
Planck 2015 results \cite{Planck}, deduced from observations of CMB
radiation anisotropies, $|\Omega_k|{<\,}0.5{\,\cdot}10^{-2}$ must
apply. From this, with use of Eq.~(\ref{eq:1.44+}) we obtain the
condition%
\begin{displaymath}
  \zeta > 8.4\,,
\end{displaymath}
which is easily complied with by our assumption $\zeta{\gtrsim}10^2$
made for Eq.~(\ref{eq:1.32}).

\subsubsection{Energy expenditure for the information transfer}

It is interesting to calculate the energy expenditure resulting from
the friction term in Eq.~(\ref{eq:3}) and used for information
transfer to newly created spatial patches. We do this in dimensionless
variables. The total positive energy of the multiverse at the
expansion $a(t)$ is%
\begin{displaymath}
  E = \varrho(a)\, V c^2
  \qquad\mbox{with}\qquad
  V = \delta\,a^3\,,
\end{displaymath}
where $V$ is the total volume and $\delta$ is a dimensionless
parameter of no further interest here. Using Eq.~(\ref{eq:1.5}a), we
now define a dimensionless energy
\begin{equation}
  \label{eq:1.46}
  \tilde{E} = \frac{E}{\varrho_\Lambda c^2l_P^3} = \rho\,\tilde{V}
  \qquad\mbox{with}\qquad
  \tilde{V} = \delta\,x^3\,.
\end{equation}
At the transition from $x$ to $x{+}\Delta x$ the energy changes by
\begin{equation}
  \label{eq:1.47}
  \Delta \tilde{E} = \delta \big(\rho'(x)\,x^3{+}3\rho(x)\,x^2\big)\Delta x
  = 3\delta \big(\gamma^2x^2{+}1/3\big)\Delta x\,,
\end{equation}
where at last Eqs.~(\ref{eq:1.5}a) and (\ref{eq:1.29}) have been used.
Simultaneously, the total energy $\tilde{E}_\Lambda{=}\delta\,x^3$
provided by $\varrho{\equiv}\varrho_\Lambda$ or $\rho{\equiv}1$
respectively changes by
\begin{equation}
  \label{eq:1.48}
  \Delta \tilde{E}_\Lambda = \Delta \tilde{V} = 3\,\delta\,x^2\Delta x\,.
\end{equation}
The energy expenditure in question is
\begin{equation}
  \label{eq:1.49}
  \Delta \tilde{E}_{ex} =  \Delta \tilde{E}_\Lambda -  \Delta \tilde{E}
  = 3\,\delta\big(x^2 {-}1/3-\gamma^2x^2\big)\Delta x
  \approx 3\,\delta\big(x^2 {-}1/3\big)\Delta x\,.
\end{equation}
(Note that in the total energy balance $\Delta \tilde{E}_{ex}$ appears
as a contribution to the negative energy of the gravitational field.)
Set in relation to the total available energy
$\Delta \tilde{E}_\Lambda$ or the energy $\Delta \tilde{E}$ supplied
to the new spatial patches, we have
\begin{equation}
  \label{eq:1.50}
   \frac{\Delta \tilde{E}_{ex}}{\Delta \tilde{E}_\Lambda} =
  1-\frac{1}{3\,x^2}
\end{equation}
or%
\vspace*{-0.5\baselineskip}
\begin{equation}
  \label{eq:1.51}
  \frac{\Delta \tilde{E}_{ex}}{\Delta \tilde{E}}
  = \frac{x^2 {-}1/3}{\gamma^2x^2{+}1/3} \approx \left\{
    \begin{array}{lll}
      3\,x^2-1\quad&\mbox{for}\quad&x\ll 1/\gamma\\
      1/\gamma^2&\mbox{for}\quad&x\gg 1/\gamma
    \end{array}
    \right.
\end{equation}
respectively. According to Eq.~(\ref{eq:1.50}),
$\Delta \tilde{E}_{ex}$ differs from $\Delta \tilde{E}_\Lambda$ for
$x{=}10$ by less than 1 percent, and with increasing x by still much
less. This means that almost all available energy is used for
information transfer over almost the entire range of x. For the energy
$\Delta \tilde{E}$, Eq.~(\ref{eq:1.51}) gives
$\Delta \tilde{E}{\approx}\Delta \tilde{E}_{ex}/2$ for $x{=}1$ and
\begin{displaymath}
  \Delta \tilde{E} \geq \gamma^2 \Delta \tilde{E}_{ex}
  \approx \gamma^2 \Delta \tilde{E}_\Lambda
\end{displaymath}
for larger $x$, where for increasing $x{\gg}1/\gamma$ the lower
boundary $\gamma^2 \Delta \tilde{E}_\Lambda$ is approached more and
more closely.

\subsubsection{Bekenstein-like bound}

It can be expected that for the storage of the information contained
in the physical laws etc. also some kind of Bekenstein bound exists.
For its form of appearance certainly plays a role, whether or not the
dark energy of the multiverse is involved in the information
storage. In the case of no involvement, our simple model provides no
possibility for any determination. We therefore turn immediately to
the case of an existing involvement.

For the reasons given in the Introduction, it makes no sense to
estimate the amount of information $I$ to be stored. Assuming,
however, that to all new space-time patches the same amount of
information will be transmitted, we can set (immediately in
dimensionless variables)%
\begin{equation}
  \label{eq:1.52}
  \Delta\tilde{I} \sim \Delta\tilde{V}\,.
\end{equation}
\mbox{}\\
\vspace*{-10mm}

\noindent
The original Bekenstein bound is $I{\leq\,}C R\,E$ where $I{=}$ stored
information, $C{=}$ constant, $R{=}$ radius of a sphere into which the
storage device fits completely, and $E{=}$ mass of the storage device
times $c^2$. Replacing $I$ with $\Delta\tilde{I}$, $R$ with $x$, $E$
with $\Delta\tilde{E}$, and using Eq.~(\ref{eq:1.52}) the Bekenstein
bound is transfered into
\begin{displaymath}
  \Delta\tilde{V} \leq \tilde{C}\,x\,\Delta\tilde{E}\,,
\end{displaymath}
where $\tilde{C}$ is an unknown number. By substituting
Eqs.~(\ref{eq:1.47}) and (\ref{eq:1.48}), from this we get
\begin{equation}
  \label{eq:1.53}
  \frac{\Delta\tilde{V}}{x\,\Delta\tilde{E}} = \frac{x}{\gamma^2x^2{+}1/3}
  \leq \tilde{C}\,.
\end{equation}
For $x{=}1/(\gamma\sqrt{3})$ the left side of this inequality assumes
a maximum with the value $\sqrt{3}/(2\gamma)$, which means, that
$\tilde{C}$ may not be smaller than this. Thus, we finally arrive at
the result that a Bekenstein-like bound of the form~(\ref{eq:1.53})
can be met, if $ \tilde{C}$ is correspondingly large, i.e.%
\begin{equation}
  \label{eq:1.54}
  \frac{\Delta\tilde{V}}{\Delta\tilde{E}}  \leq \frac{\sqrt{3}\,x}{2\gamma}\,.
\end{equation}
As discussed in paragraph~(\ref{su:3}) of Section~\ref{sec:suppl},
there is still some flexibility with respect to the parameter $\gamma$
so that even a lower upper limit could be met.%
\mbox{}\\
\vspace*{1.5mm}

\hspace*{1.5mm}A similar transfer of the Bremermann limit does not seem
possible because the volume of the device for information storage
would have to be constant over time. But that does not change the fact
that, as assumed in the Introduction, there will certainly exist a
speed limit to the information transfer considered by us.

\newpage

\section{\small Interpreting \large$x(\tau)$ \small as a
  solution of the FL-equation for\\ a scalar field \large $\varphi$}
\label{sec:Interpret}

In Ref.~\cite{Rebhan_1} it was shown, that to any prescribed density
$\varrho(x)$ with $\varrho'(x){\leq}0$ (a condition necessitated by
Eq.~(\ref{eq:2.1}) below and shown to be satisfied by our present
results at the end of Subsection~\ref{sec:varrho}) functions
$\varphi(x)$, $v(\varphi)$ and $x(\tau)$ exist such that
Eqs.~(\ref{eq:1.10})-(\ref{eq:1.12}) are satisfied. To show that in
the present case these are meaningful solutions, we derive them now
explicitly for the density~(\ref{eq:1.29}). Here, Eq.~(23b) of
Ref.~\cite{Rebhan_1} assumes the form%
\begin{equation}
  \label{eq:2.1} 
  \dot{\varphi}(\tau) = \sqrt{-\frac{x\,\varrho'(x)}{3\,\varrho_\Lambda}}
\end{equation}
and can be obtained by differentiating Eq.~(\ref{eq:1.11}a) with
respect to $\tau$ and inserting Eq.~(\ref{eq:1.12}). Combining
Eqs.~(\ref{eq:1.10}) and (\ref{eq:2.1}) we get
\begin{equation}
  \label{eq:2.2}
  \frac{d\varphi}{dx} = \frac{\dot{\varphi}(\tau)}{\dot{x}(\tau)}
  =\sqrt{-\frac{x\,\varrho'(x)}{3\,(\varrho\,x^2{-}\varrho_\Lambda)}}\,,
\end{equation}
while resolving Eq.~(\ref{eq:1.11}a) with respect to $v$ and inserting
the square of Eq.~(\ref{eq:2.1}) yields%
\begin{equation}
  \label{eq:2.3}
  v = \frac{\varrho}{\varrho_\Lambda} +
  \frac{x\,\varrho'(x)}{6\,\varrho_\Lambda}\,.
\end{equation}
Inserting Eq.~(\ref{eq:1.29}) in Eqs.~(\ref{eq:2.2}) and
(\ref{eq:2.3}), and using Eq.~(\ref{eq:1.32}) gives
\begin{displaymath}
  \frac{d\varphi}{dx}  =  \frac{1}{x^2}\,\sqrt{\frac{2\,\varrho_\Lambda}{3\,\varrho_0}}
  \,,\qquad
  v = \frac{\varrho_0}{\varrho_\Lambda} + \frac{2}{3\,x^2}
\end{displaymath}
and
\vspace*{-0.5\baselineskip}
\begin{equation}
  \label{eq:2.4}
  \varphi(x) = \sqrt{\frac{2\,\varrho_\Lambda}{3\,\varrho_0}}\,
  \left(1-\frac{1}{x}\right) =
  \sqrt{\frac{2\,\varrho_\Lambda}{3\,\varrho_0}}\,\big(1{-}\e^{-\gamma\tau}\big)\,.
\end{equation}
Thereby, an integration constant was chosen such that $\varphi(1){=}0$,
and at last Eq.~(\ref{eq:1.24}) was used. Using
$h{=}\dot{x}(\tau)/x{=}\gamma$ according to Eq.~(\ref{eq:1.24}), from
this we get
\begin{displaymath}
  \left|\frac{\ddot{\varphi}(\tau)}{3h\,\dot{\varphi}(\tau)}\right|
  = \frac{1}{3}\,,
\end{displaymath}
which means, that $\varphi(x(\tau))$ is no slow roll solution.

Resolving Eq.~(\ref{eq:2.4}) with respect to $x$ and inserting the
function $x(\varphi)$ thus obtained in our last result for $v$, with
use of Eq.~(\ref{eq:1.32}) we finally obtain
\begin{equation}
  \label{eq:2.5}
  v(\varphi) = \frac{\varrho_0}{\varrho_\Lambda} +
  \frac{2}{3}\,
  \left(\sqrt{\frac{3\,\varrho_0}{2\,\varrho_\Lambda}}\,\varphi-1\right)^2
  = \gamma^2 +  \frac{2}{3}\, \left(\sqrt{\frac{3}{2}}\,\gamma\varphi-1\right)^2\,.
\end{equation}
This is a parabolic potential assuming the extremely small minimum
$v_{min}{=}\gamma^2$ at $\varphi{=}\sqrt{2/3}/\gamma$. According to
Eq.~(\ref{eq:2.4}), $\varphi$ starts with the value~$0$ at the time
$\tau{=}0$ and reaches the minimum only after an infinite time,
$\tau{\to}\infty$. This is due to $v_{min}{\neq}0$ because in a
parabolic potential with vanishing minimum, $\varphi$ would reach it
after a finite time and would oscillate around it. It is very
appropriate that such oscillations are avoided because they would
spoil our multiverse concept. Note that in our model $v_{min}{\neq}0$
is no arbitrary assumption but comes about inevitably through the
underlying concept.

\section{Supplements}
\label{sec:suppl}

\begin{enumerate}

\item\label{su:1} Let us return for a moment to the effect which the
  inclusion of matter with mass density $\varrho_m$ would have
  according to Ref.~\cite{Rebhan_1}. Here, too, by an appropriate mix
  of the initial densities $\varrho$ of dark energy and $\varrho_m$ of
  matter it could be achieved, that the initial state is an
  equilibrium -- albeit an unstable one \cite{Rebhan_4}. The multiverse
  could thus remain still longer close to the latter than it would do
  according to FIGs.~\ref{fig:1}-\ref{fig:3}. When it finally moves
  away, $\varrho_m{\sim}x^{-n}$ becomes, very soon, much smaller than
  $\varrho{\sim}x^{-2}$ (see Eq.~(\ref{eq:1.33});
  $\varrho_m/\varrho{\leq}10^{-2}$ for $x{\geq}10$ if $n{=}4$ and for
  $x{\geq}100$ if $n{=}3$). For the largest part of the evolution
  $\varrho_m$ can therefore be neglected.

\item\label{su:2} During the whole lifetime of our universe, i.e. for
  $\mathcal{T}{-}0.98{\leq}\mathcal{T}{\leq}\mathcal{T}_0$, the
  equation $\varrho{=}\varrho_0{=}const$ applies with high
  precision. Furthermore, by choosing $\zeta$ sufficiently large,
  according to Eq.~(\ref{eq:1.44+}) the spatial curvature can be made
  so small that within our universe there are no measurable
  differences to flat space. Added together, this leads to an
  evolution of our universe just as in the cosmological standard model
  with small cosmological constant. The proof of this is not trivial
  considering the differences in the underlying space-time. However,
  because there are no relevant differences from the case considered
  in Ref.~\cite{Rebhan_1}, it suffices to refer to the proof given
  there in Section~3.2. The same holds for the repercussion of our
  universe on the multiverse (there Section~3.3).

\item\label{su:3} The value ${\approx}10^{122}$ of the ratio
  $\varrho_\Lambda/\varrho_0$ used in this paper is not unalterably
  fixed but could be adjusted to other theoretical values within
  several orders of magnitude. This can be achieved by imposing the
  initial conditions~(\ref{eq:1.7}), required for the continuous
  connection to a tunneling process, at a value
  $a_i{\gg}l_P$. Choosing for example $a_i{=}10\,l_P$ or
  $a_i{=}10^{16}l_P$, according to Eq.~(\ref{eq:1.8}) this would yield
  $\varrho_\Lambda/\varrho_0{\approx}10^{120}$ or
  $\varrho_\Lambda/\varrho_0{\approx}10^{90}$ respectively. Here, the
  initial state of the multiverse would still be far in the quantum
  regime, and the approximate treatment of the initial tunneling
  process by continuation of the classical solution for $\tau{\geq}0$
  to imaginary values $u{=}\mbox{i}\,\tau$ for $\tau{\leq}0$ is then
  justified for an even wider range of $u$~values. A further aspect
  may call for $a_i{\gg}l_P$: For encoding all information about the
  physical laws, similar to the case of the DNA, a certain minimum
  initial volume $V{=}2\pi^2 a_i^3$ could be required which for
  $a_i{=}l_P$ might be too small.

\item\label{su:4} The ansatz~(\ref{eq:3}) can be extended to contain
  nonlinear friction terms such as ${-}g\dot{a}^2(t)$ with
  constant $g$. The disadvantage involved is a non-linearity of the
  differential equation to be solved. An expansion with respect to $g$
  would offer itself for handling this problem.

It can also be shown, that solutions $x(\tau)$, $\varphi(\tau)$ of
Eqs.~(\ref{eq:1.10})-(\ref{eq:1.12}) of the standard theory with given
potential $v(\varphi)$ satisfy Eq.~(\ref{eq:1.13}) with
$2\sigma\dot{x}(\tau)$ replaced by $f(x)\,\dot{x}(\tau)$, i.e.%
\vspace*{-0.5\baselineskip}
\begin{equation}
  \label{eq:3.2}
  \ddot{x}(\tau) + f(x)\,\dot{x}(\tau) -x = 0\,.
\end{equation}
From this equation we get
\vspace*{-0.2\baselineskip}
\begin{equation}
  \label{eq:3.3}
  \dot{x}^2(\tau) = x^2 - 2\!\int_0^\tau \!\!f(x(\tau'))\,\dot{x}^2(\tau')\,d\tau'
\end{equation}
in the same way as Eqs.~(\ref{eq:4})-(\ref{eq:5}) are obtained from
Eq.~(\ref{eq:3}). This equation must coincide with Eq.~(\ref{eq:1.10})
why we must have
\begin{displaymath}
  2\!\int_0^\tau\!\!\! f\big (x(\tau')\big )\dot{x}^2(\tau')\,d\tau' =
  -\left(\frac{\varrho}{\varrho_\Lambda}{-}1\right)x^2\,.
\end{displaymath}
Differentiating this equation with respect to $\tau$, dividing it by
$\dot{x}(\tau)$ and resolving it with respect to $f(x)$ yields
\begin{equation}
  \label{eq:3.4}
  f(x) = -\frac{x\varrho'\!(x)/2{+}\varrho(x){-}\varrho_\Lambda}
  {\varrho_\Lambda\sqrt{\varrho(x)/\varrho_\Lambda{-}1/x^2}}
  = -\frac{x\rho'\!(x)/2{+}\rho(x){-}1}
  {\sqrt{\rho(x){-}1/x^2}}\,.
\end{equation}
Thereby we used Eq.~(\ref{eq:1.5}a), (\ref{eq:1.10}) and
$\dot{\varrho}(\tau){=}\varrho'(x)\,\dot{x}(\tau)$. The function
$\varrho(x)$ is obtained from Eq.~(\ref{eq:1.11}a) by inserting in it
$\tau{=}\tau(x)$, the inverse function of the solution $x(\tau)$.  By
resolving Eq.~(\ref{eq:1.10}) or $x^2\rho(x){=}1{+}\dot{x}^2(\tau)$
for $x$ and inserting the result $x{=}g(\dot{x}) $ in $f(x)$,
Eq.~(\ref{eq:3.2}) can even be brought into the form
\begin{equation}
  \label{eq:3.5}
  \ddot{x}(\tau) + h\!\left(\dot{x}(\tau)\right) -x = 0
  \qquad\mbox{with}\qquad
  h\!\left(\dot{x}(\tau)\right) =
  \dot{x}(\tau)f\!\left(g\left(\dot{x}\right)\right)\,.
\end{equation}

\item\label{su:5} The results of Section~\ref{sec:Interpret} enable a
  covariant generalization of Eq.~(\ref{eq:4}) with (\ref{eq:5}) and
  thus also indirectly of the ansatz~(\ref{eq:3}). The covariant
  contribution of $\varrho_\Lambda$ to the field equations is
  $\Lambda g_{\mu\nu}$, where $\Lambda{=}8\pi G\varrho_\Lambda/c^2$
  (see e.g. p.~394 of Ref.~\cite{Rebhan_3}). Concerning the total
  density $\varrho{=}\varrho_f{+}\varrho_\Lambda$ or rather
  $\rho{=}\varrho_f/\varrho_\Lambda{+}1$, we take advantage of the
  fact that, using Eq.~(\ref{eq:1.11}a), it can be represented by the
  scalar field $\varphi(x)$ of Eq.~(\ref{eq:2.4}) and the potential
  $v(\varphi)$ of Eq.~(\ref{eq:2.5}). (Eq.~(\ref{eq:1.13}) follows
  from this by use of Eqs.~(\ref{eq:3.4}), (\ref{eq:3.2}),
  (\ref{eq:1.31}) and (\ref{eq:1.23}).)  In the Einstein field
  equations a (dimensioned) scalar field $\Phi$ is covariantly
  represented by the energy-momentum tensor (see e.g. p. 527 of
  Ref.~\cite{Weinberg_2} or p. 549 of Ref.~\cite{Rebhan_3})
\begin{displaymath}
  T_{\mu\nu} = \frac{\hbar}{\mu}(\partial_\mu\Phi)(\partial_\nu\Phi)
  -\left[\frac{\hbar^2}{2\mu}(\partial^\lambda\Phi)(\partial_\lambda\Phi)-V(\Phi)
  \right] g_{\mu\nu}\,.
\end{displaymath}
Under the symmetries of the cosmological principle and in
Robertson-Walker coordinates, from this follows%
\vspace*{-0.3\baselineskip}
\begin{displaymath}
  \label{eq:3.6}
  \frac{T_{00}}{c^2} = \varrho\,,
\end{displaymath}
where $\varrho$ is given by Eq.~(\ref{eq:1.1}a). With this,
$g_{00}{=}1$ according to the Robertson-Walker metric, and the
definition $T^{(f)}_{00}/c^2{=}\varrho_f$ we can write
$\varrho{=}\varrho_f{+}\varrho_\Lambda$ in the form
\begin{equation}
  \label{eq:3.7}
  \frac{T_{00}}{c^2} = \frac{T^{(f)}_{00}}{c^2} + \varrho_\Lambda g_{00}\,. 
\end{equation}
Now we define%
\vspace*{-0.3\baselineskip}
\begin{equation}
  \label{eq:3.8}
  T^{(f)}_{\mu\nu} = T_{\mu\nu} - \varrho_\Lambda c^2\,g_{\mu\nu}\,. 
\end{equation}
Because $T_{\mu\nu} $ and $\varrho_\Lambda c^2\,g_{\mu\nu}$ are
tensors, this is also true for their difference $T^{(f)}_{\mu\nu}$.
Since under given circumstances Eq.~(\ref{eq:3.8}) reduces to
Eq.~(\ref{eq:3.7}), it is the covariant generalization of the latter.
The corresponding covariant field equations also include a
generalization of Eq.~(\ref{eq:3}), although the kind of
representation is different. (In an analog representation, the place
of $\dot{a}(t)$ should be taken by $g_{\mu\nu;\lambda}$ rather than by
an expression involving $\Phi$.)  It is, however, questionable whether
under lower symmetries the corresponding field equations allow for a
similar simple interpretation as that given by Eq.~(\ref{eq:1.13}),
(\ref{eq:3.2}), or (\ref{eq:3.5}).

\end{enumerate}

\section{Conclusions}

As already indicated in the Introduction, the main ideas of this paper
are speculative. This means, that they may be far away from reality,
contain at least some truth, or constitute a step in the right
direction. In any case, the solutions following from them turn out
particularly well suited for generating an all-embracing multiverse
which provides the dark energy of our universe and thus explains its
presence there. At the end of the Introduction and in
Sec.~\ref{sec:Interpret} it was shown that they are also valid
solutions of the cosmological standard theory and may be considered as
a re-interpretation of the latter. What is thus a vindication on the
one hand is a disadvantage for detecting the value of the underlying
concept on the other, since the solutions themselves are not
appropriate for a discrimination between the two interpretations. To
accept the conception of a dark energy field with huge mass
density~$\varrho_\Lambda$, increasingly weakened by some agent --
information transfer in our model~--, may thus be considered as a
question of taste.

The above-mentioned special usefulness of our new solutions for the
purposes pursued in Ref.~\cite{Rebhan_1} was not foreseeable at first,
because their mathematical structure is already completely determined
through the new conceptual assumptions. Within the standard theory
they could not be derived without some rather artificial assumptions,
what may be regarded as some support to the newly added ideas. The
solutions are very simple, and through the relation
$\gamma{=}\sqrt{\varrho_0/\varrho_\Lambda}$ they depend in a very
transparent way on the enormous discrepancy between $\varrho_0$ and
$\varrho_\Lambda$ on which our new concept is based. Furthermore, they
are easy to handle in spite of their dependence on the extremely small
parameter $\gamma$ which allows for the very simple and nevertheless
exceedingly good approximation~(\ref{eq:1.24}). In the framework of
the cosmological standard theory they avoid oscillations around the
necessarily non-vanishing minimum of their quadratic potential
$v(\varphi)$. During the whole lifetime of our universe the equation
$w{=}p/(\varrho c^2){=}{-}1$ is almost exactly satisfied and thus
provides the best possible fit to observational data. The reason is,
that, in spite of marked differences in the distant past, the
solutions cling from today's perspective already for quite some time
very tightly to those of a cosmological constant model with
$\varrho{=}\varrho_0$.

In contrast to the cosmological standard model assuming flat space,
our model employs a finite space with positive curvature. The latter
can be chosen so small that within our universe there are no
measurable differences. Furthermore, during the lifetime of our
universe the changes of the dark energy density $\varrho$ are
immeasurably small. Therefore, our model provides the same internal
evolution of our universe as the cosmological standard model with
cosmological constant. On the other hand, further outside the
observable region, there is a significant difference: While there is
no reason for spatial expansion in the causally disconnected external
regions of flat space, the corresponding regions in the positively
curved space of our model experience a strong and continuously
increasing spatial expansion. Aside from that, as in the models of
Ref.~\cite{Rebhan_1} the dark energy density $\varrho$ is a monotonic
function of the spatial curvature and is different from zero at all
finite times.  In consequence, the space-time spanned by the
multiverse is firmly bound to the presence of matter or energy and
thus fulfills an important consequence of Mach's principle \cite{Mach},
a fact which does not apply to a flat-space multiverse.

\end{document}